\documentclass[twoside,fleqn]{ActaStyle}
\usepackage{times,cite}

\usepackage{lscape}             % if you need landscape enviroinment
\usepackage{graphicx,epsfig}    % if you need to include figures in PostScript
\usepackage[tbtags]{amsmath}    % if you need more math

\def\bge{\begin{equation}}
\def\ene{\end{equation}}
\def\bg{\begin{eqnarray}}
\def\en{\end{eqnarray}}

\def\ubar{{\bar{u}}}
\def\dbar{{\bar{d}}}
\def\sbar{{\bar{s}}}

\def\vr{\vec{r}}

\def\authorheading{Steven D. Bass}

\def\shorttitle{Gluonic effects in $\eta$ and $\eta'$ nucleon and nucleus
interactions}

\begin{document}
%% ---------------------------------------------------

%%%  page range, first and last page
\pagerange{1}{6}

%%% paper title
\title{%
GLUONIC EFFECTS IN
$\eta$- and $\eta'$-NUCLEON AND NUCLEUS INTERACTIONS
}

%%% author(s) and address(es)
\author{%  author(s)
Steven D. Bass
}
{%  address(es)
Institute for Theoretical Physics, University of Innsbruck \\
Technikerstrasse 25, A6020 Innsbruck, 
Austria }

%%% Date of submition
\day{November 23, 2005}

\abstract{
Gluonic degrees of freedom play an important role in the masses of 
the $\eta$ and $\eta'$ mesons. 
We discuss $\eta$- and $\eta'$-nucleon and nucleus interactions 
where this glue may be manifest.
Interesting processes being studied in experiments are $\eta'$
production in proton-nucleon collisions close to threshold and
possible $\eta-$nucleus bound-states.}


\section{The axial U(1) problem}

Gluonic degrees of freedom play an important role in the physics of the 
flavour-singlet $J^P = 1^+$ channel \cite{uppsala} through the QCD axial 
anomaly \cite{zuoz}.
The most famous example is the axial U(1) problem: the masses of the 
$\eta$ and $\eta'$ mesons are much greater than the values they 
would have if these mesons were pure Goldstone bosons associated
with spontaneously broken chiral symmetry \cite{weinberg}.
This extra mass is induced by non-perturbative gluon dynamics 
\cite{thooft,hfpm,gvua1,witten,vecca,ks}.

Spontaneous chiral symmetry breaking is associated with a non-vanishing
chiral condensate
\begin{equation}
\langle \ {\rm vac} \ | \ {\bar q} q \ | \ {\rm vac} \ \rangle < 0
.
\label{eq5}
\end{equation}
The non-vanishing chiral condensate also spontaneously breaks the axial
U(1) symmetry so, naively, we expect a nonet of would-be pseudoscalar 
Goldstone bosons:
the octet associated with chiral $SU(3)_L \otimes SU(3)_R$
plus a singlet boson associated with axial U(1)
--- each with mass squared 
$m^2_{\rm Goldstone} \sim m_q$ where $m_q$ denotes the light and 
strange quark masses.
The pions and kaons are described well by this theory.
The masses of the $\eta$ and $\eta'$ mesons
are
about 300-400 MeV too big to fit in this picture without additional 
physics.
One needs extra mass in the singlet channel associated 
with
non-perturbative gluon configurations and
the QCD axial anomaly \cite{zuoz}.
The strange quark mass induces considerable $\eta$-$\eta'$ mixing.
For free mesons
the $\eta - \eta'$ mass matrix (at leading order in the chiral expansion) 
is
\begin{equation}
M^2 =
\left(\begin{array}{cc}
{4 \over 3} m_{\rm K}^2 - {1 \over 3} m_{\pi}^2  &
- {2 \over 3} \sqrt{2} (m_{\rm K}^2 - m_{\pi}^2) \\
\\
- {2 \over 3} \sqrt{2} (m_{\rm K}^2 - m_{\pi}^2) &
[ {2 \over 3} m_{\rm K}^2 + {1 \over 3} m_{\pi}^2 + {\tilde m}^2_{\eta_0} ]
\end{array}\right)
.
\label{eq10}
\end{equation}
Here ${\tilde m}^2_{\eta_0}$ denotes the gluonic mass contribution
in the singlet channel.
It has a rigorous interpretation through the Witten-Veneziano mass 
formula
\cite{witten,vecca}
and
is associated with non-perturbative gluon topology, related perhaps 
to confinement \cite{ks} or instantons \cite{thooft}.
When we diagonalize this matrix
\begin{eqnarray}
| \eta \rangle &=&
\cos \theta \ | \eta_8 \rangle - \sin \theta \ | \eta_0 \rangle
\\ \nonumber
| \eta' \rangle &=&
\sin \theta \ | \eta_8 \rangle + \cos \theta \ | \eta_0 \rangle
\label{eq11}
\end{eqnarray}
with
\begin{equation}
\eta_0 = \frac{1}{\sqrt{3}}\; (u\ubar + d\dbar + s\sbar),\quad
\eta_8 = \frac{1}{\sqrt{6}}\; (u\ubar + d\dbar - 2 s\sbar)
\label{mixing2}
\end{equation}
we obtain values for the $\eta$ and $\eta'$ masses
\begin{equation}
m^2_{\eta', \eta}
= (m_{\rm K}^2 + {\tilde m}_{\eta_0}^2 /2)
\pm {1 \over 2}
\sqrt{(2 m_{\rm K}^2 - 2 m_{\pi}^2 - {1 \over 3} {\tilde m}_{\eta_0}^2)^2
   + {8 \over 9} {\tilde m}_{\eta_0}^4}
.
\label{eq12}
\end{equation}
The physical mass of the $\eta$ is close to the octet mass
$
m_{\eta_8} = \sqrt{ {4 \over 3} m_{\rm K}^2 - {1 \over 3} m_{\pi}^2 }
$,
within a few percent.
However, to build a theory of the $\eta$ treating it as a pure octet
state
risks losing essential physics associated with the singlet component
and axial U(1) dynamics.
In the absence of the gluonic term (${\tilde m}_{\eta_0}^2$ set equal
to zero), 
one finds
$m_{\eta'} \sim \sqrt{2 m_{\rm K}^2 - m_{\pi}^2}$
and
$m_{\eta} \sim m_{\pi}$.
That is, without extra input from glue, in the OZI limit,
the $\eta$ would be approximately an isosinglet light-quark state
(${1 \over \sqrt{2}} | {\bar u} u + {\bar d} d \rangle$)
degenerate with the pion and
the $\eta'$ would be a strange-quark state $| {\bar s} s \rangle$
--- mirroring the isoscalar vector $\omega$ and $\phi$ mesons.
The gluonic mass term is vital to understanding the physical $\eta$ 
and $\eta'$ mesons.
\footnote{
Taking the value ${\tilde m}_{\eta_0}^2 = 0.73$GeV$^2$ 
\cite{vecca}
in the leading-order mass formula, Eq.(\ref{eq12}),
gives agreement with the physical masses at the 10\% level.
The
corresponding
$\eta - \eta'$
mixing angle $\theta \simeq - 18^\circ$
is within the range $-17^\circ$ to $-20^\circ$ obtained
from a study of various decay processes in \cite{gilman,frere}.
Closer agreement with the physical masses can be obtained
by introducing
the singlet decay constant
$F_0 \neq F_{\pi}$ and including higher-order mass terms in the chiral
expansion
\cite{leutwyler,feldmann}.}

\section{Glue and $\eta$ and $\eta'$ nucleon interactions}

Given that glue plays an important role in the masses of the $\eta$ and 
$\eta'$ mesons, it is worthwhile and interesting to look for possible 
manifestations of gluonic effects in dynamical processes involving these 
mesons.
In the rest of this paper we consider $\eta$ and $\eta'$ production in 
proton-nucleon collisions close to threshold, and possible $\eta$--nucleus 
bound-states.
These systems are being studied in experiments at COSY and GSI.
We note that the $\eta'$--nucleon coupling constant
is related, in part, 
to the flavour-singlet axial-charge extracted 
from polarized deep inelastic scattering experiments 
\cite{shorev}
-- for a recent review see \cite{spin}.

\subsection{$\eta$ and $\eta'$ production in proton-nucleon collisions
close to threshold}

Since the singlet components of the $\eta$ and $\eta'$ couple to glue, 
it is natural to consider the process where glue is excited in the 
``short distance'' ($\sim 0.2$fm) 
interaction region of a proton-nucleon collision and then evolves to 
become an $\eta'$ in the final state \cite{bass99}.
This gluonic induced production mechanism is extra to the contributions 
associated with 
meson exchange models \cite{holinde,wilkin,faldt}.
Given the large gluonic effect in the mass, 
there is no reason, a priori, to expect it to be small.
The contribution to the matrix elements for $\eta'$ and $\eta$ 
production is weighted
by the singlet-component projection-factors
$\cos \theta$ for the $\eta'$ and $\sin \theta$ for the $\eta$
where $\theta$ 
is 
the $\eta - \eta'$ mixing angle.
The angle $\theta \sim -20$ degrees means that gluonic induced 
production should 
be considerably enhanced in $\eta'$ production compared to $\eta$ production.

What is the phenomenology of this gluonic interaction ?

Since glue is flavour-blind the gluonic production process has the same 
size in both the $pp \rightarrow pp \eta'$ and $pn \rightarrow pn \eta'$ 
reactions.
CELSIUS \cite{celsius} have measured the ratio
$R_{\eta} 
 = \sigma (pn \rightarrow pn \eta ) / \sigma (pp \rightarrow pp \eta )$
for quasifree $\eta$ 
production from a deuteron target up to 100 MeV above threshold.
They observed that $R_{\eta}$ is approximately energy-independent 
$\simeq 6.5$ over the whole energy range.
The value of this ratio signifies a strong isovector exchange 
contribution to the $\eta$ production mechanism \cite{celsius}.
This experiment is being repeated for $\eta'$ production.
The cross-section for $pp \rightarrow pp \eta'$ 
close to threshold has been measured by the COSY-11 Collaboration
\cite{cosy}
who are now
measuring the $pn \rightarrow pn \eta'$ process \cite{cosyprop}.
In the extreme scenario that the glue-induced production saturated 
the $\eta'$ 
production cross-section, the ratio
$R_{\eta'} =
 \sigma (pn \rightarrow pn \eta' ) / \sigma (pp \rightarrow pp \eta' )$
would go to one
after we correct for the final state interaction \cite{faldt,protonfsi}
between the two outgoing nucleons.
In practice, we should expect contributions from both gluonic 
and meson-exchange type mechanisms.
It will be interesting to observe the ratio $R_{\eta'}$ and how it compares 
with $R_{\eta}$.

Gluonic induced production appears as a contact term 
in the axial U(1) extended chiral Lagrangian for low-energy QCD \cite{bass99}.

\subsection{$\eta$--nucleus bound-states}

New experiments at the GSI will employ the recoilless $(d, \ ^3He)$
reaction to study the possible formation of $\eta$ meson bound states
inside the nucleus \cite{hayano,gillitzer},
following on from the successful studies of pionic atoms in these reactions
\cite{pionexpt}.
The idea is to measure the excitation-energy spectrum and then, if a clear
bound state is observed,
to extract the in-medium effective mass, $m_{\eta}^*$, of the $\eta$
in nuclei through performing a fit to this spectrum with the $\eta$-nucleus
optical potential.

Meson masses in nuclei are determined from the scalar induced contribution
to the meson propagator evaluated at zero three-momentum, ${\vec k} =0$, in
the nuclear medium.
Let $k=(E,{\vec k})$ and $m$ denote the four-momentum and mass of the meson 
in free space.
Then, one solves the equation
\begin{equation}
k^2 - m^2 = {\tt Re} \ \Pi (E, {\vec k}, \rho)
\end{equation}
for ${\vec k}=0$
where $\Pi$ is the in-medium $s$-wave meson self-energy and $\rho$ is the 
nuclear density.
Contributions to the in medium mass come from coupling to the scalar
$\sigma$ field in the nucleus in mean-field approximation,
nucleon-hole and resonance-hole excitations in the medium.
The $s$-wave self-energy can be written as \cite{ericson}
\begin{equation}
\Pi (E, {\vec k}, \rho) \bigg|_{\{{\vec k}=0\}}
=
- 4 \pi \rho \biggl( { b \over 1 + b \langle {1 \over r} \rangle } \biggr) .
\end{equation}
Here 
$
b = a ( 1 + {m \over M} )
$
where
$a$ is the meson-nucleon scattering length, $M$ is the nucleon mass and
the mean inter-nucleon
seperation is $\langle {1 \over r} \rangle$.
Attraction corresponds to positive values of $a$.
The denominator in Eq.(7) is the Ericson-Ericson double scattering correction.

The in-medium mass $m_{\eta}^*$ is sensitive to the flavour-singlet
component in the $\eta$, and hence 
to the non-perturbative glue associated with axial U(1) dynamics.
An important source of the in-medium mass modification comes
from light-quarks
coupling to the scalar $\sigma$ mean-field in the nucleus.
Increasing the flavour-singlet component in the $\eta$
at the expense of the octet component gives more attraction,
more binding and a larger value of the $\eta$-nucleon
scattering length, $a_{\eta N}$.
Since the mass shift is approximately proportional to the $\eta$--nucleon 
scattering length, it follows that that the physical value of $a_{\eta N}$ 
should be larger than if the $\eta$ were a pure octet state.

This physics has been investigated by Bass and Thomas \cite{bt05}.
QCD arguments suggest that the gluonic mass term is suppressed at 
finite density due to coupling to the $\sigma$ mean-field in the nucleus.
\footnote{
In the chiral limit the singlet
analogy to the Weinberg-Tomozawa
term does not vanish because of the anomalous glue terms.
Starting from the simple Born term one finds
anomalous gluonic contributions
to the singlet-meson nucleon scattering length
proportional to ${\tilde m}^2_{\eta_0}$ and ${\tilde m}_{\eta_0}^4$
\cite{bassww}.
}
Phenomenology is used
to estimate the size of the effect in the $\eta$
using
the Quark Meson Coupling model (QMC) of hadron properties in the nuclear 
medium
\cite{etaqmc}.
Here one uses the large $\eta$ mass
(which in QCD is induced by mixing and the gluonic mass term)
to motivate taking an MIT Bag
description
for the $\eta$ wavefunction, and
then coupling the light (up and down)
quark and antiquark fields in the $\eta$ to the scalar $\sigma$
field
in the nucleus working in mean-field approximation \cite{etaqmc}.
The strange-quark component of the wavefunction does not couple
to the $\sigma$ field.
$\eta-\eta'$ mixing is readily built into the model.

The mass for the $\eta$ in nuclear matter is self-consistently
calculated by solving for the MIT Bag in the nuclear medium \cite{etaqmc}:
\begin{equation}
m_\eta^*(\vr) = \frac{2 [a_P^2\Omega_q^*(\vr)
+ b_P^2\Omega_s(\vr)] - z_\eta}{R_\eta^*}
+ {4\over 3}\pi R_\eta^{* 3} B,
\label{meta}\\
\end{equation}
\begin{equation}
\left.\frac{\partial m_j^*(\vr)}
{\partial R_j}\right|_{R_j = R_j^*} = 0, \quad\quad (j = \eta, \ \eta').
\label{equil}\\
\end{equation}
Here $\Omega_q^*$ and $\Omega_s$ are light-quark and strange-quark
Bag energy eigenvalues,
$R_{\eta}^*$ is the Bag radius in the medium and $B$ is the Bag constant.
The $\eta-\eta'$ mixing angle $\theta$ is included in the terms
$
a_P = \frac{1}{\sqrt{3}} \cos\theta
- \sqrt{\frac{2}{3}} \sin\theta
$
and
$
b_P = \sqrt{\frac{2}{3}} \cos\theta
+ \frac{1}{\sqrt{3}} \sin\theta
$
and can be varied in the model.
One first solves the Bag for the free $\eta$ with a
given mixing angle, and then turns on QMC to obtain the mass-shift.
In Eq.~(\ref{meta}), $z_\eta$ parameterizes the sum of 
the center-of-mass and gluon fluctuation effects, and is 
assumed to be independent of density~\cite{finite0}.

The coupling constants in the model for the coupling of light-quarks
to the $\sigma$ mean-field in the nucleus are adjusted to fit 
the saturation energy and density of symmetric nuclear matter and the 
bulk symmetry energy.
The Bag parameters used in these calculations are
$\Omega_q = 2.05$
(for the light quarks)
and $\Omega_s = 2.5$
(for the strange quark)
with
$B = ( 170 {\rm MeV} )^4$.
For nuclear matter density we find $\Omega^*_q = 1.81$ for the 1$s$ state.
This value depends on the coupling of light-quarks to the $\sigma$ mean-field
and is independent of the mixing angle $\theta$.

Increasing the mixing angle increases the amount of singlet
relative to octet components in the $\eta$.
This produces greater attraction through increasing
the amount of light-quark compared to strange-quark
components in the $\eta$
and a reduced effective mass.
Through Eq.(7) increasing the mixing angle
also increases the
$\eta$-nucleon scattering length $a_{\eta N}$.
We quantify this in Table 1 which presents results
for the pure octet
($\eta=\eta_8$, $\theta=0$)
and the values
$\theta = - 10^\circ$ and $- 20^\circ$ (the physical mixing angle).
%,
%%%%%%%%%%%%%%%%%%%%%%%%%%%%%%%%%%%%%%%%%%%%%%%%%%%%%%%%%%%%%%%%%%%%%%%
\begin{table}[htbp]
\begin{center}
\caption{
Physical masses fitted in free space, 
the bag
masses in medium at normal nuclear-matter
density,
$\rho_0 = 0.15$ fm$^{-3}$,
and the corresponding meson-nucleon scattering lengths.
}
\label{bagparam}
\begin{tabular}[t]{c|lcll}
\hline
&$m$ (MeV) & $m^*$ (MeV) & ${\tt Re} a$ (fm)
\\
\hline
$\eta_8$  &547.75  & 500.0 &  0.43 \\
$\eta$ (-10$^o$)& 547.75  & 474.7 & 0.64 \\
$\eta$ (-20$^o$)& 547.75  & 449.3 & 0.85 \\
$\eta_0$  &      958 & 878.6  & 0.99 \\
$\eta'$ (-10$^o$)&958 & 899.2 & 0.74 \\
$\eta'$ (-20$^o$)&958 & 921.3 & 0.47 \\
\hline
\end{tabular}
\end{center}
\end{table}
The values of ${\tt Re} a_{\eta}$ quoted in Table 1 are obtained
from substituting the in-medium and free masses into Eq.~(7) with
the Ericson-Ericson denominator turned-off, and using the free
mass in the expression for $b$. The effect of exchanging $m$ for
$m^*$ in $b$ is a 5\% increase in the quoted scattering length.
The QMC model makes no claim about the imaginary part of the scattering
length.
The key observation is that $\eta - \eta'$ mixing leads to a factor of
two increase in the mass-shift of the $\eta$ meson and in 
the scattering length obtained in the model.
\footnote{
Because the QMC model has been explored mainly at the mean-field level, it is
not clear that one should include the Ericson-Ericson term in extracting the
corresponding $\eta$ nucleon scattering length.
Substituting the scattering lengths given in Table 1 into Eq.~(7)
(and neglecting the imaginary part)
yields resummed values
$a_{eff} = a / ( 1 + b  \langle 1/r \rangle )$
equal to 0.44 fm for the $\eta$
with the physical mixing angle $\theta = -20$ degrees, with corresponding 
reduction in the binding energy.
}

The density dependence of the mass-shifts in the QMC model is discussed
in Ref.\cite{etaqmc}.
Neglecting the Ericson-Ericson term, the mass-shift is approximately
linear. For densities $\rho$ between 0.5 and 1 times $\rho_0$ (nuclear
matter density) we find
\begin{equation}
m^*_{\eta} / m_{\eta} \simeq 1 - 0.17 \rho / \rho_0
\end{equation}
for the physical mixing angle $-20^\circ$.
The scattering lengths extracted from this analysis are density independent 
to within a few percent over the same range of densities.

\section{Conclusions and Outlook}

Glue plays an important role in the masses of the $\eta$ and $\eta'$
mesons. 
New experiments are measuring the interactions of these mesons 
with nucleons and nuclei. 
The glue which generates a large part of the $\eta$ and $\eta'$ 
masses can contribute to the cross-section 
for $\eta'$ production in proton-nucleon collisions and 
to the possible binding energies of $\eta$ and $\eta'$ mesons in nuclei.
It will be interesting to see the forthcoming data from COSY and GSI on
these processes.

\begin{ack}
The work of SDB is supported by the Austrian Research Fund, FWF,
through contract P17778.
\end{ack}

\end{document}